%% file: main.tex
\documentclass[letterpaper,onecolumn,11pt]{article}
\usepackage[margin=1in]{geometry}
\usepackage{epsfig,endnotes}
\usepackage{hyperref}
\usepackage{url}
\usepackage{amssymb,amsfonts,amsmath,amsthm,amscd,dsfont,mathrsfs}
\usepackage{graphicx,float,psfrag,epsfig,enumitem}
\usepackage{wrapfig}
\usepackage{color}

\usepackage{xspace}
\usepackage{booktabs} 
\usepackage{cleveref}
\usepackage{bbm,tikz}

\usepackage[utf8]{inputenc}
\usepackage[utf8]{inputenc} 
\usepackage[T1]{fontenc}    
\usepackage{hyperref}       
\usepackage{url}            
\usepackage{booktabs}       
\usepackage{amsfonts}       
\usepackage{nicefrac}       
\usepackage{microtype}      
\usepackage{bbm}
\usepackage{wrapfig}
\usepackage{multirow}
\usepackage{subcaption,prettyref}
\usepackage{graphicx}
\usepackage{color}
\usepackage{makecell}
\usepackage{etoolbox}

\usepackage{makecell}
\usepackage{etoolbox}

\usepackage{bm}
\usepackage{bbm}
\usepackage{tcolorbox}

\usepackage{float}
\usepackage{xcolor}

\usepackage{algorithm}
\usepackage[noend]{algpseudocode}
\usepackage{algorithmicx}
\algnewcommand{\LeftComment}[1]{\Statex {\color{teal}\textbf{\(\triangleright\) #1}}} 
\algdef{SxnE}[IF]{As}{EndAs}[1]{\textbf{as}\ #1\ }
\algdef{SxnE}[IF]{Upon}{EndUpon}[1]{\textbf{upon}\ #1\ \algorithmicdo}

\makeatletter
\newsavebox{\@brx}
\newcommand{\llangle}[1][]{\savebox{\@brx}{\(\m@th{#1\langle}\)}%
  \mathopen{\copy\@brx\kern-0.5\wd\@brx\usebox{\@brx}}}
\newcommand{\rrangle}[1][]{\savebox{\@brx}{\(\m@th{#1\rangle}\)}%
  \mathclose{\copy\@brx\kern-0.5\wd\@brx\usebox{\@brx}}}
\makeatother

\def\<{\langle}
\def\>{\rangle}

\newcommand{\lr}[1]{\langle #1 \rangle}

\newcommand{\prepare}{\textsc{proposal}\xspace}
\newcommand{\vone}{\textsc{vote}\xspace}
\newcommand{\fs}{\text{Forensic Storage}\xspace}
\newcommand{\detector}{\text{Detector}\xspace}
\newcommand{\latestqc}{\textit{highQC}\xspace}
\newcommand{\lockqc}{\textit{lockQC}\xspace}
\newcommand{\currentround}{\textit{curRound}\xspace}
\newcommand{\parentqc}{\textit{parentQC}\xspace}
\newcommand{\parenthash}{\textit{parentHash}\xspace}
\newcommand{\timeout}{\textsc{timeout}\xspace}

\title{Accountability and Forensics in Blockchains:\\ XDC Consensus Engine DPoS 2.0}

\author{Gerui Wang, Jerome Wang, Liam Lai, Fisher Yu\\
\textsf{Hash Laboratories}\\ \textsf{\{gerui.wang, jerome.wang, liam.lai, fisher.yu\}@hashlabs.cc}}

\date{August 2021}

\begin{document}

\maketitle

\begin{abstract}
This document introduces XinFin DPoS 2.0, the proposed next generation decentralized consensus engine for the XinFin XDC Network. Built upon the most advanced BFT consensus protocol, this upgrade will empower the XDC Network with military-grade security and performance while consuming extremely low resources, and will be fully backwards-compatible in terms of APIs. It will also pave the road to the future evolution of the XDC Network.

The core invention is the holistic integration of {\em accountability and forensics} in blockchains: the ability to identify malicious actors with cryptographic integrity directly from the blockchain records,  incorporating the latest peer-reviewed academic research \cite{sheng2020bft} with  state of the art engineering designs and implementation plans. 

\end{abstract}

\newpage
\tableofcontents
\newpage

\input{intro}

\input{section2}

\input{section3}

\input{section4}

\input{acknowledgement}

\bibliographystyle{plain}
\bibliography{reference.bib}

\end{document}

%% file: intro.tex
\section{Introduction}
Trust systems are a core construct of human societies.  For centuries, mutually distrustful parties have collaborated to build empires, economies, and social structures. These collaborations and interactions, however, have historically been managed with opaque systems that are susceptible to corruption and extreme imbalance of power. 

{\em Decentralized trust systems} are digital systems in which multiple parties can collaborate on specific tasks without requiring parties to trust one another. An example of a task that requires decentralized trust is that of running a financial system, including payments. Decentralized trust systems promise panacea from the ills of centralization and associated corruptibility.  Although decentralized computer systems have traditionally posed substantial technical challenges (e.g., scalability and security in peer-to-peer networks), substantial breakthroughs in blockchain technologies in recent years have paved the way for the mainstream development and deployment of decentralized trust systems: 

 {\em Timeline of Blockchain developments}. 
 \begin{itemize}
     \item {\sf Blockchain 1.0}:  Bitcoin and proof of work (PoW) \cite{nakamoto2019bitcoin}. Highly  secure protocol and system design with guaranteed consensus as long as majority of hash power is honest. The major drawback is that the system is {\em very inefficient}. The inefficiencies are in {\em energy} as well as in {\em scalability}: poor throughput (only a handful of transactions per second) and poor latency (hour of confirmation time). {\em Summary}: This era resulted in the first construction of a permissionless blockchain that provides an immutable ledger with extremely strong security properties and resistance to malicious actors, only now being fully explored. 
     \item  {\sf Blockchain 2.0}: Ethereum and programmable smart contracts  \cite{wood2014ethereum}. This design maintains the  security of Bitcoin (and its scalability deficiencies) but adds a general purpose programmable platform, allowing a large family of applications. {\em Summary}: This era resulted in the invention of EVM (Ethereum Virtual Machine), the ``blockchain computer", allowing general purpose software programs to be implemented atop the blockchain. 
     \item {\sf Blockchain 3.0}: {\em Proof of Stake} (PoS). The PoS protocols  eschew the energy consuming PoW mining process and simultaneously allow high throughput and low  latency (e.g., Ouroboros \cite{kiayias2017ouroboros}, SnowWhite \cite{daian2019snow}, Algorand \cite{gilad2017algorand}, Hotstuff  \cite{yin2019hotstuff}, Tendermint \cite{kwon2014tendermint}). The PoS  blockchain protocols have  been constructed from a family of  protocols  guaranteed to provide consensus even when a fraction of participants act maliciously (so-called Byzantine behavior) - known as BFT (Byzantine Fault Tolerant) protocols \cite{lamport2019byzantine} in Computer Science.  {\em Summary}: This era focused on  the scaling problem of  blockchains, improving throughput and latency by orders of magnitudes. 
 \end{itemize}
 
 {\em Security of Blockchains}. 
  A defining feature of blockchains is its security against adversarial actions, the so-called Byzantine fault tolerance. A standard format of expressing this {\em security} property is the following: 
\begin{quote}
as long as a fraction $x$\% of participation level is {\em honest}, i.e., follows protocol, then the security of the blockchain is guaranteed.    
\end{quote}
In PoW blockchains like Bitcoin, the participation level is measured in {\em mining hash power} and the security threshold $x$ is 0.5. In PoS blockchains like Algorand, the participation level is measured in {\em staking power} and security threshold $x = \frac{1}{3}$. In a BFT protocol such as Hotstuff, the participation level is measured in {\em number of permissioned users} and security threshold $x = \frac{1}{3}$. When the Byzantine participation level crosses the security threshold of $x$,  security is not guaranteed: in fact, such a scenario represents a doomsday scenario and ``all bets are off". 

\begin{figure}[htb]
    \centering
    \includegraphics[width=0.6\textwidth]{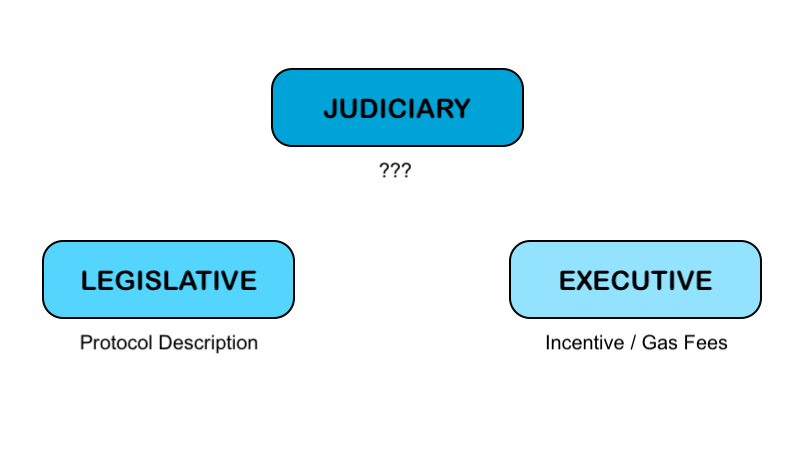}
    \caption{Judiciary, a core component of governance, is missing in extant blockchain designs. Addressing this lacuna in a systematic manner represents the core  goal of this whitepaper.}
    \label{fig:democracy}
\end{figure}

An alternative view point of this state of affairs is provided by anthropological studies of human governance systems. Over millennia, stable human governance systems have evolved to a three-party architecture: legislative,  executive and judiciary wings. Indeed, this architecture is the dominant format in all extant governing systems. In the context of blockchains, the legislative aspect is clearly codified via the description of the protocol (the longest chain protocol in {\sf Bitcoin}) and participation strategies (e.g., proof of work mining in {\sf Bitcoin}). The executive branch is represented by the incentive (e.g., block rewards in {\sf Bitcoin}) and taxing mechanisms (e.g., transaction fees (in {\sf Bitcoin} and gas fees (in {\sf Ethereum}). A striking  aspect in all extant blockchain designs is the  absence of a judiciary system; see Figure~\ref{fig:democracy}. Addressing this lacuna is the core goal of this whitepaper. 

{\sf Blockchain 4.0}: {\em Accountability and Attributability}. Trust underlies human societies and key aspects of incentivizing and maintaining trustworthy behavior are the notions of accountability (viewed internally) and attibutability (imposed  externally). As an example, self-reporting financial statements  (accountability) and auditing (attribution) are standard checks and balances of all financial systems. Key to mainstream blockchain adoption and usage, especially in the financial services sector, is to build accountability to blockchains. This notion of financial accountability was introduced in blockchains via the notion of {\em slashing conditions} in \cite{buterin2017casper} - however, this protocol was not completely specified and has since been overtaken by other versions \cite{buterin2020combining} in the development of {\sf Ethereum 2.0}. In particular, no  extant blockchain, either in production or design stage incorporates forensics holistically (one notable exception is \cite{crain2021red}, which aims to design new blockchains).
We propose that the next era of blockchain designs will be focused on this key property, with XDC network spearheading this revolution,  summarized as follows:

\begin{center}

\fbox{%
    \parbox{5in}{%
    {\sf XDPoS 2.0}  {\bf Credo}: The blockchain  is secure when sufficient participation is honest; conversely if security is breached, then  every malicious actor is identified (with  cryptographic integrity).  
}
}
\end{center}

In  this whitepaper, we detail the design of such a blockchain; we build on very recent academic research  \cite{sheng2020bft} in pursuing this design process.

%% file: section2.tex
\section{XDPoS 2.0}

Sitting at the core of the XDC Network, XinFin Delegated Proof-of-Stake consensus protocol 2.0 (XDPoS 2.0) regulates the XDC nodes in maintaining the consistency of a decentralized ledger (the blockchain) with strong security and performance guarantees. In this section, we provide an overview of XDPoS 2.0 by introducing its three pillars, including:
\begin{enumerate}
    \item Master node election, which specifies how the delegation and proof-of-stake works.
    \item The consensus engine, namely, the HotStuff state machine replication (SMR) protocol, which is the state-of-the-art BFT (Byzantine fault-tolerant) SMR (state-machine replication) protocol. An illustration of its position in XDPoS 2.0 and a brief security analysis will be provided.
    \item Reward mechanism, which incentivizes nodes to join and maintain the XDC Network.
\end{enumerate}

\subsection{Master Node Election}
\label{sec:21}
Unlike proof-of-work (PoW) which wastes resources and has low performance, XDPoS 2.0 uses Delegated Proof-of-Stake to decide which nodes have the right to create the next block and how the created block is approved. XinFin nodes that contribute in creating and approving blocks are referred to as {\it master nodes}. Only XinFin nodes that meet certain stake and hardware criteria are eligible as the master nodes. The eligible criteria are as follows:
\begin{itemize}
    \item More than 10,000,000 XDC deposited into the XinFin smart contract.
\item A suitable wallet to store XDC tokens. Preferably in hardware form.
\item A dedicated and stable hardware environment.
\item A dedicated Static Public IP address.
\item 100\% network uptime by IDC network.
\item A minimum of tier 3+ IDC environment.
\item Virtual Private Server (VPS). Though this optional, this option is highly recommended.
\item When using cloud-based services like Amazon EC2 M3, large virtual machine (VM) sizes are appropriate. Similar configurations are applicable for the Microsoft Azure Cloud network users.
\end{itemize}

Once a node meets these criteria, it becomes a master node candidate and is eligible for master node elections that happen periodically. More specifically, time is partitioned into epochs. At the beginning of each epoch, a random set of 108 master candidates are elected as the master nodes of this epoch. These master nodes forms a BFT committee that is responsible for creating 900 consecutive blocks of the chain (the delegation). Every block is certificated by the super-majority of the committee (and thus is finalized) following the Hotstuff protocol. Sample aforementioned parameter values are:
\[\texttt{VALIDATOR\_SET\_SIZE=108};\qquad \texttt{EPOCH=900}.\]

The random election can be achieved via various mechanisms:
\begin{enumerate}
    \item Rank candidates by deposit and elect the top 108 candidates.
    \item Using verifiable random functions (VRFs) to randomly elect 108 candidates. The probability of being elected should be proportional to deposit.
\end{enumerate}
XDPoS 2.0 will start with mechanism (1) for backwards compatibility, and can switch to mechanism (2) seamlessly for added fairness and security.

Once the election is done, all nodes that are not elected will keep listening to messages sent by the HotStuff BFT committee and passively maintain the blockchain. We now introduce the HotStuff protocol.

\subsection{The HotStuff Protocol}
\label{sec:22}
\subsubsection{Protocol Description}
HotStuff is a state-of-the-art Byzantine fault-tolerant (BFT) state machine replication (SMR) protocol widely used as an enterprise grade blockchain consensus engine, such as Facebook's Novi and Diem (previously known as Libra) projects. Compared to the Nakamoto longest chain (BTC, ETH, etc), it has two distinct advantages:

\begin{enumerate}
    \item Deterministic security (a.k.a. finality), which means zero forking / rollback on confirmed blocks as long as the fraction of adversary master nodes is less than 1/3. In contrast, Nakamoto only provides probabilistic security, that is, there is a non-zero chance of rollback.
    \item Resiliency to network partial asynchrony, which means that HotStuff remains secure even if the network is temporarily not synchronous (e.g., temporary large latency between different sets of nodes). In contrast, Nakamoto is insecure under network partial asynchrony.
\end{enumerate}


\paragraph{Setup} 
HotStuff aims at reaching consensus within a BFT committee, which is a deterministic set of nodes that know each other (recognizing their IP address, account address, and public key, etc.). In XDPoS 2.0, this BFT committee is the set of master nodes determined at the beginning of each epoch. Furthermore, these master nodes are ordered (e.g. by ascending order of their account address) to facilitate the protocol.

The consensus to be reached in XDPoS 2.0 is about an ever-growing list of blocks. Each block has a parent hash pointer. Thus, together they form a blockchain. The transitive closure of the parent relation is called ancestors.

The protocol is a leader-based one proceeding in rounds. Each round has a different leader, which is chosen among the ordered master nodes in a  round-robin manner. In other words, the leader of round $r$ ($0\leqslant r < EPOCH$) is the master node whose index is $(r\mod \texttt{VALIDATOR\_SET\_SIZE})$. The role of a leader is to propose a new block and collect votes from the remaining nodes.



\paragraph{Data Structure}

\begin{figure}[htb]
    \centering
    \includegraphics[width=0.6\textwidth]{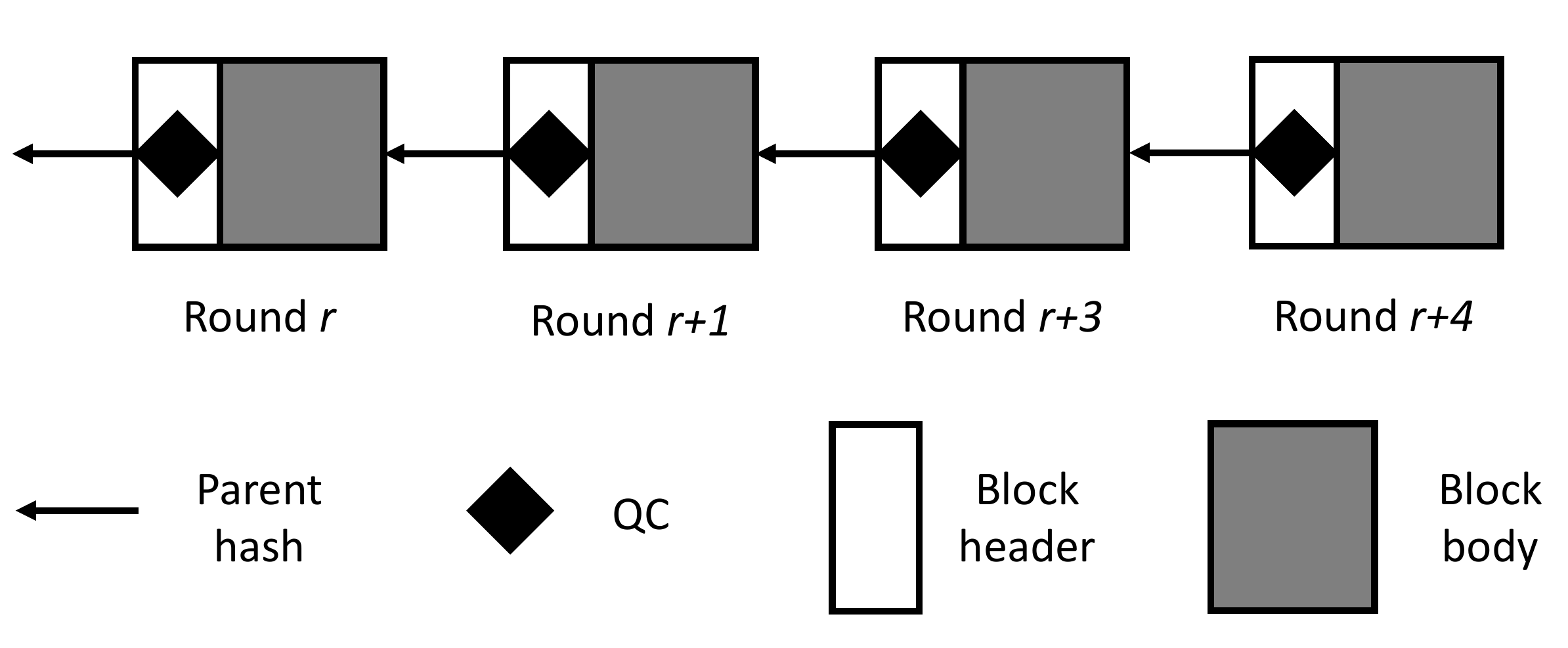}
    \caption{XDC blockchain}
    \label{fig:hotstuff-chain}
\end{figure}

The block data structure is similar to that in Ethereum, except that a \parentqc field is added to the block header. Here QC stands for quorum certificate. It is created by the leader of round-$r$ for the block in $r-1$ (the parent block) when at least $t_H=\left \lceil{\texttt{VALIDATOR\_SET\_SIZE}\times 2/3}\right \rceil$ master nodes have voted for this parent block, certifying that this parent block has been approved by the super-majority of the nodes. A QC contains the parent block hash\footnote{This parent hash is the same as header.\parenthash. But we do not override header.\parenthash for the sake of backwards-compatibility.}, the round number, and the metadata such as signatures of the vote messages. In case of no proposal and/or a timeout in a round $r-1$, nodes will send a "timeout" message for this round, and the leader of round $r$ will gather $t_H$ of them into a timeout certificate (TC), so that this round can be dropped from the blockchain. We note that round number is not the same as block number - round number might be skipped in the blockchain but block numbers are guaranteed to be consecutive. Figure~\ref{fig:hotstuff-chain} shows an example of the main chain with QCs. Note that round $r+2$ is skipped in the main chain due to timeout. The specification of data structures is provided in Algorithm~\ref{alg:data-structure}.

\begin{algorithm}[htb]
\caption{Data Structures in the protocol}
\label{alg:data-structure}
\begin{algorithmic}[0]
    \State block
    \State \hspace{\algorithmicindent}$\parentqc$: QC for the parent block, stored in header
    \State \hspace{\algorithmicindent}$parent$: short for $\parentqc.block$
    \State \hspace{\algorithmicindent}$\ldots$: other fields are the same as XDPoS 1.0
    \item[]
    \State QC
    \State \hspace{\algorithmicindent}$round$: the round of the voted block
    \State \hspace{\algorithmicindent}$block$: the hash of the voted block
    \State \hspace{\algorithmicindent}$signatures$: signatures of votes
    \item[]
    \State TC
    \State \hspace{\algorithmicindent}$round$: the round of the timeout
    \State \hspace{\algorithmicindent}$signatures$: signatures of timeout
\end{algorithmic}
\end{algorithm}



\paragraph{Proposing and Voting}
Each round of HotStuff follows the proposing-voting paradigm: In each round $r$, the leader proposes a block and all master nodes vote. The leader of round $r+1$ will collect these votes and aggregate them into a QC if there are $t_H$ votes. The leader of round $r+1$ then proposes a child block of the block proposed in round $r$, and add the QC of the parent block to its header. Master nodes should obey the following rules regarding QC and TC:
\begin{enumerate}
    \item \textbf{Leader proposal.} When a master node is the leader of a round, it must propose a new block where the \parentqc field carries the latest QC it has received. 
    \item \textbf{The locking rule.} When a master node receives a valid proposal of a block with a \parentqc, it \textit{locks} on the \parentqc contained in the parent block of the proposal, i.e., it \textit{locks} on the grandparent block of the proposal. Once it locks on a block, it will ignore all the blocks that are not the offspring of this block, except under certain conditions (see the voting rule). The locking rule contributes to the guarantee that the HotStuff protocol has zero forking in finalized blocks/blockchain.
    \item \textbf{The voting rule.} When a master node receives a proposal of a block, it votes  for it only if it already locks on this block or this block's ancestor. The only exception happens when the round of this proposal block's parent block is higher than the block it locks on.
    \item \textbf{The finalization rule.} Define a 3-chain as a tuple of 3 blocks $[b,b',b'']$ such that $b=b'.$parent and $b'=b''.$parent.  When a node (either master or non-master node) receives a 3-chain of blocks in three consecutive rounds and an extra QC whose \parenthash is the hash of $b''$, it finalizes the first block $b$ and all its ancestors.
    \item \textbf{Time out handling.} If a master node does not receive any proposal for the current round within a certain time period, it must send a timeout message for this round. Sufficient (2/3 of the nodes) timeout messages can form a TC that helps master nodes to move on to the next round.
\end{enumerate}

The protocol is presented in Algorithm~\ref{alg:hotstuff}. Notice that we slightly abuse the notation by referring $qc.block$ to the actual block instead of the block hash when the context is clear. And we omit the operations of verifying signatures of signed messages and verifying metadata in QC for simplicity.

\begin{algorithm}[!thb]
\caption{HotStuff protocol}
\label{alg:hotstuff}
\begin{algorithmic}[1]
    \State $t_H\gets\left \lceil{\texttt{VALIDATOR\_SET\_SIZE}\times 2/3}\right \rceil$\Comment{constant}
    \State $\latestqc\gets qc_{genesis}$; $\lockqc\gets qc_{genesis}$ \Comment{the latest QC and lock variable}
    \LeftComment{Optimistic path}
    \For{$\currentround\gets1,2,\ldots$}
    \As{a leader}
        \State generate proposal block $b^*$ containing $\latestqc$ and block content (transactions, etc.)
        \State broadcast $\lr{\prepare,\currentround,b^*}$
    \EndAs{}
    \As{a master node}
        \State on receiving $\lr{\prepare,\currentround,b^*}$ from a leader
        \label{alg:check-leader-1}
        \If{$b^*$ satisfies the voting rule}
            \State send $\lr{\vone,\currentround,H(b^*)}$ to the next leader \Comment{$H()$ means the hash digest}
        \EndIf
        \State $\latestqc\gets \max_{round}\{b^*.\parentqc, \latestqc\}$\Comment{don't update in case of a draw}
    \State $\lockqc\gets \max_{round}\{b^*.\parentqc.block.\parentqc, \lockqc\}$\Comment{don't update in case of a draw}
    \EndAs{}
    \As{the next leader}
        \If{collect $\lr{\vone,\currentround,h^*}$ from $t_H$ committee members for the same hash $h^*$}
        \State $\latestqc\gets $ QC generated from those \vone messages
        \EndIf
    \EndAs{}
    \EndFor
    \LeftComment{At any time in round $\currentround$, as either master or non-master nodes}
    \Upon{receiving $qc\gets$ QC in \prepare message}
    \State $b''\gets qc.block$, $b'\gets b''.parent$, $b\gets b'.parent$ 
    \If{$b,b',b''$ are in consecutive rounds}
    \State finalize blocks through $b$, execute transactions in the finalized blocks
    \EndIf
    \If{$qc.round>\currentround$}
    \State $\currentround\gets qc.round+1$ and go to the beginning of $\currentround$
    \EndIf
    \EndUpon
    
    \LeftComment{Non-optimistic path}
    
    \Upon{a local timeout}
    \State stop voting for $\currentround$
    \State broadcast $\lr{\timeout,\currentround}$ along with $\latestqc$
    \If{enter $\currentround$ due to a TC $tc$}
    \State broadcast $tc$
    \EndIf
    \EndUpon
    \Upon{receiving $t_H$ \timeout messages for the same round}\label{alg:check-voter-2}
    \State generate TC $tc$ from them and broadcast $tc$
    \EndUpon
    \Upon{seeing a TC $tc$}
    \If{$tc.round>\currentround$}
    \State $\currentround\gets tc.round+1$ and go to the beginning of $\currentround$
    \EndIf
    \EndUpon
\end{algorithmic}
\end{algorithm}

\subsubsection{Safety and Liveness Guarantee}
\paragraph{Safety}
Safety means any two honest nodes should agree on (finalize) the same XDC Network blockchain. And the HotStuff protocol ensures safety in the following two ways: safety within a round and safety across rounds.

The HotStuff protocol requires that an honest master node only votes once in any round. Therefore, when the Byzantine adversary does not corrupt more than 1/3 master nodes, there cannot be two conflicting QCs within a round (a.k.a. no equivocation). This is because a valid QC requires 2/3 votes. The existence of two valid QC mean that more than 1/3 master nodes have double voted, which is not possible since it requires some honest master nodes to double vote together with all the 1/3 adversary nodes.

When the adversary does not corrupt more than 1/3 master nodes, HotStuff ensures safety across rounds by the use of \textit{locks} and the \textit{voting rule}. If a QC for block $b$ exists in round $r$, at least 2/3 master nodes lock on $(r,b)$. In rounds higher than $r$, there could not be a QC on a conflicting block whose ancestors don't include $b$, since such a QC means more than 1/3 master nodes who lock on $(r,b)$ have voted for the conflicting block. This, in turn, means more than 1/3 master nodes violate the voting rule, which contradicts with our assumption of 1/3 adversary.

\paragraph{Liveness} 

Liveness means the blockchain makes progress and new transactions are included. When the leader is honest, the adversary does not corrupt more than 1/3 master nodes, and the network communication is synchronous, the HotStuff protocol ensures that a valid block with transactions can be generated. In addition, whenever this condition happens at four consecutive rounds (3-chain and an extra QC), the first two blocks are finalized.

\subsubsection{Performance Guarantees}
The HotStuff protocol is optimistically responsive: in optimistic condition (honest leader and synchronous network), it only takes 3 block arrival time (BAT) to finalize a block and the transactions it contains. In practice, due to our strict network requirement on master nodes eligibility, the maximum network round-trip latency should be well below 2 seconds. Therefore, we can set BAT to be 2 seconds. This means that, in optimistic conditions, the finalization latency is only 6 seconds.

In terms of overhead, the size of QC is only a few kilo bytes (for example, using the ETH 65-byte signature, a QC from 2/3 master nodes of a 108-node committee is only $65 * 2/3 * 108 = 4.68$kB). This is much smaller than the common block size (e.g., ETH block size is about 60kB and is growing) and thus is negligible. Moreover, HotStuff's communication complexity per round is linear to the product of committee size and block size, which is order-optimal.

\subsection{Reward Mechanism}
\label{sec:23}

One distinct feature of XDPoS 2.0 is that the blocks are finalized quickly, allowing the reward to be determined and announced instantly after the block is finalized. Since the optimal block finalization latency is 6 seconds, master nodes can expect to receive the reward in 6 seconds. Apart from the reduced reward latency, the reward mechanism remains the same as in XDPoS 1.0.

%% file: section3.tex
\section{What to Expect}
\subsection{Seamless Upgrade}
\begin{figure}[htb]
    \centering
    \includegraphics[width=0.9\textwidth]{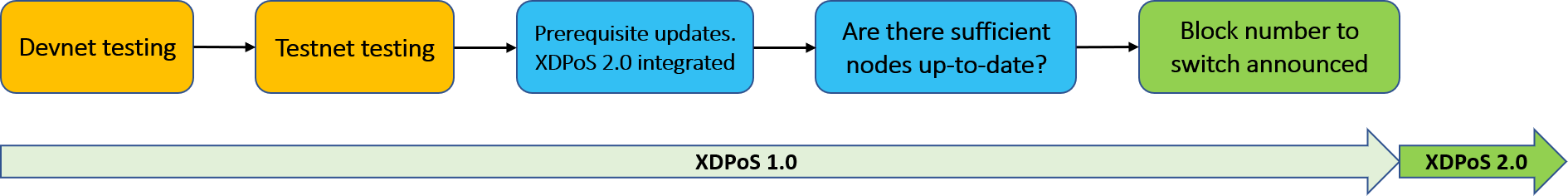}
    \caption{XDPoS 2.0 upgrade plan}
    \label{fig:upgrade_plan}
\end{figure}

A seamless upgrade is our top priority. Before the upgrade, our QA and product team will thoroughly test the new consensus engine for at least 3 months in lab Devnet and public testnet. Once it is proved bulletproof, the team will release a series of prerequisite updates to integrate an inactive version of XDPoS 2.0 into the system. Once a sufficient number of nodes have updated, we will announce the block number after which XDPoS 2.0 will be activated and only blocks created using XDPoS 2.0 will be accepted. This enables a seamless upgrade without any interruptions to the liveness of the system.

The upgrade will be fully backwards-compatible in terms of data and APIs. After the upgrade, XDC network will continue to extend the current public ledger from the 1.0 era. All the existing wallet addresses, XDCs, stakes, transactions, and ledger states will be intact and can be used exactly the same as before. The APIs to wallet, browser, exchanges, etc. will be fully backwards compatible as well. Therefore, for XinFin customers, the upgrade will be transparent.

\subsection{Military Grade Security and Performance}
As we have proved in the last section, XinFin DPoS 2.0 guarantees absolute zero forking in the finalized blockchain as long as the ratio of adversarial master nodes in each epoch does not exceed 1/3. No network communication issues could cause forking, either.

Performance wise, no resource-intensive mining will be involved at all. The block time will still be 2 seconds and the throughput will remain the same. The average transaction confirmation latency will be 3 BATs, i.e., 6 seconds after the inclusion of the transaction in a block. Most importantly, after the upgrade, confirmation also means finalization. In other words, transactions can be finalized in merely 6 seconds after its inclusion in a block.

\subsection{New Feature: Forensic Monitoring}
We are also excited to introduce a new feature of XinFin system enabled by the consensus engine upgrade, namely forensic monitoring.

In the unlikely case where the ratio of Byzantine nodes exceeds 1/3, it is possible for these nodes to collude and create a security violation --- either a safety violation or a liveness violation. However, they cannot achieve this without being held accountable and punished, thanks to the protocol design.


\input{forensics_short}

%% file: forensics_short.tex
\paragraph{Safety Violation and Forensic Monitoring}
When the adversary corrupts more than 1/3 master nodes in the BFT committee of an epoch, it is possible for the adversary to violate the safety and jeopardize the consensus by creating forks - such as two finalized blockchains. However, certain messages need to be signed and sent by these nodes to make this happen, which can be detected by the system and served as irrefutable evidence of the misbehavior. Those messages are embedded into the blockchain and can be obtained by querying master nodes for forked blockchains.

This property of XDPoS 2.0 enables our safety forensic feature, which can identify as many Byzantine master nodes as possible while obtaining the proof from querying as few witnesses as possible. The process of identifying culpable Byzantine master nodes involve performing appropriate quorum intersections: since two quorums of $t_H\approx \texttt{VALIDATOR\_SET\_SIZE}\times 2/3$ master nodes intersect in at least $\texttt{VALIDATOR\_SET\_SIZE}/3$ master nodes, we are able to identify that many Byzantine master nodes, which is the optimal number we can identify. As for the witnesses, two honest nodes having access to one of the two conflicting blockchains, respectively, are sufficient for the proof. Since non-master nodes also passively monitor the blockchain, they can serve as witnesses as well. This also implies that as long as there are at least two honest nodes in the XDC network and the adversary wants to create a safety violation for the two nodes, we can provide this forensic feature. 

Once a master node is held culpable by the forensic protocol, the proof should be provided to a XinFin governance-driven penalty mechanism (e.g. a slashing smart contract). 

\paragraph{Liveness Violation and Forensic Monitoring}
In XDPoS 2.0, Byzantine master nodes can also slow down the blockchain instead of creating forks. This behavior is called underperforming, and can take three forms:
\begin{enumerate}
    \item a master node fails to propose a block when it is the leader;
    \item the leader proposes a block but does not use the latest QC as \parentqc;
    \item a master node (non-leader) fails to propagate vote messages.
\end{enumerate}

If there are more than 1/3 underperforming master nodes, the performance of XDPoS 2.0 could be decreased and we call it a liveness violation.

However, unlike safety violation, there is no cryptographic evidence for liveness violation to hold underperforming master nodes culpable. The handling of liveness violation is thus softer than safety violation: Master nodes should broadcast a \textit{blame} message if they believe another master node is underperforming. 
Those messages are collected and reported to XinFin governance-driven penalty mechanism for final decision, such as excluding the suspect node from master node election for certain number of epochs or raising its deposit requirement.


\begin{figure}[!htb]
    \centering
    \includegraphics[width=0.8\columnwidth]{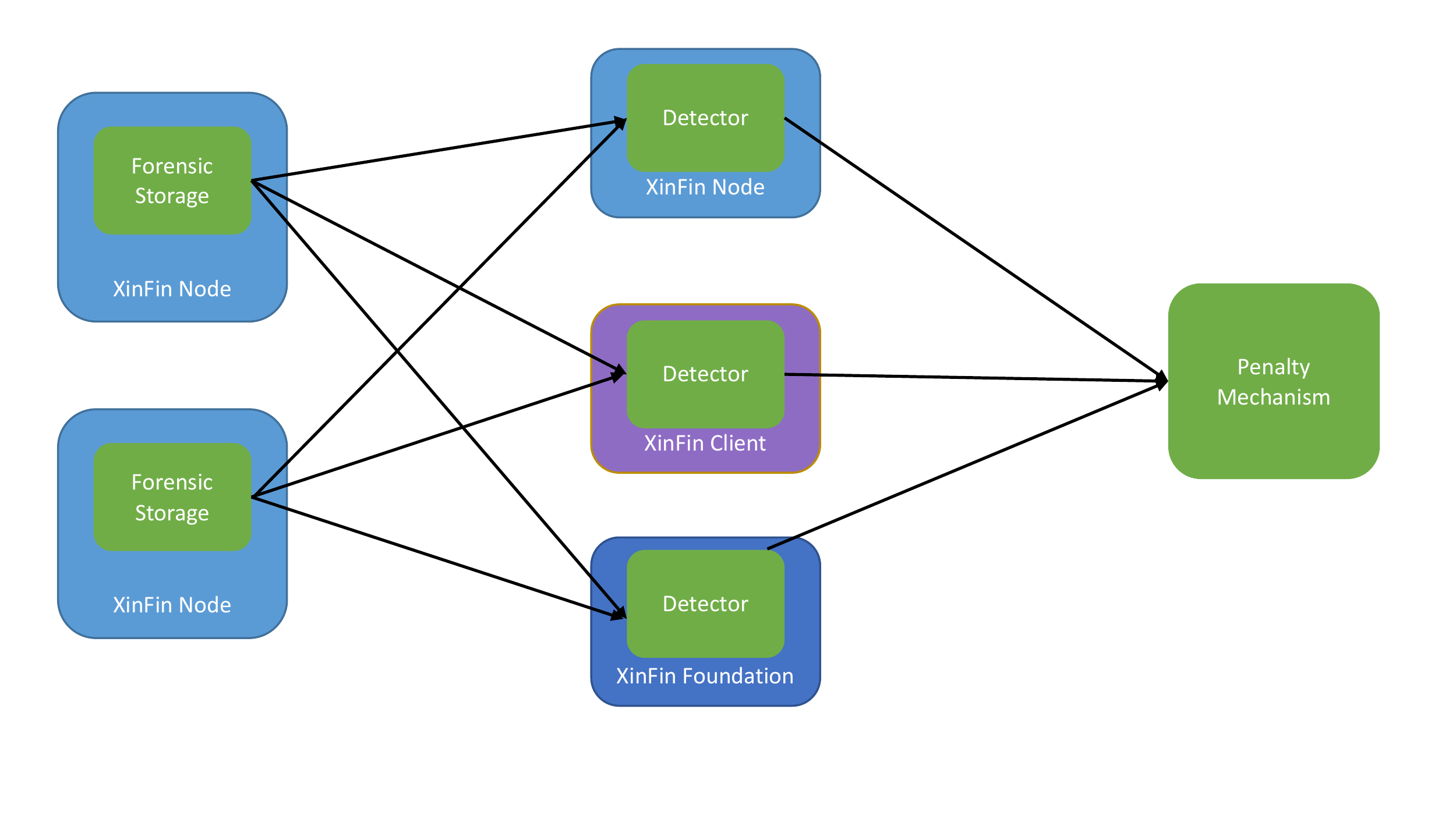}
    \caption{Forensic module structure and information flow.}
    \label{fig:forensic_store}
\end{figure}

\paragraph{Implementation}
We plan to implement the corresponding forensic monitoring protocol as a module on top of the HotStuff protocol. 
The module consists of three components and this structure is shown in Figure~\ref{fig:forensic_store}.

\begin{itemize}
    \item \textbf{\fs}, a database at the XinFin nodes that stores forensic information. It maintains a map from the round number to quorum certificates, blocks, and their persistent storage. In addition, it records the blame messages as well. It can be accessed by other participants of the system, including other nodes (e.g., via RPC requests).  
    \item \textbf{\detector}, which can be run by any participant of XinFin system. It sends requests periodically to connected nodes and use the returned information for safety and liveness forensic monitoring. If it finds enough information to penalize a master node, it will send the information to the penalty mechanism.
    \item \textbf{Penalty Mechanism}, a XinFin governance-driven body that decides the penalty for safety and liveness violation. It implements both hard penalty (e.g. slashing Byzantine nodes' deposits via a slashing smart contract or excluding them from master node selection) and soft penalty (e.g. raising the deposit requirement). 
\end{itemize}

\begin{figure}[!htb]
    \centering
    \includegraphics[width=0.8\textwidth]{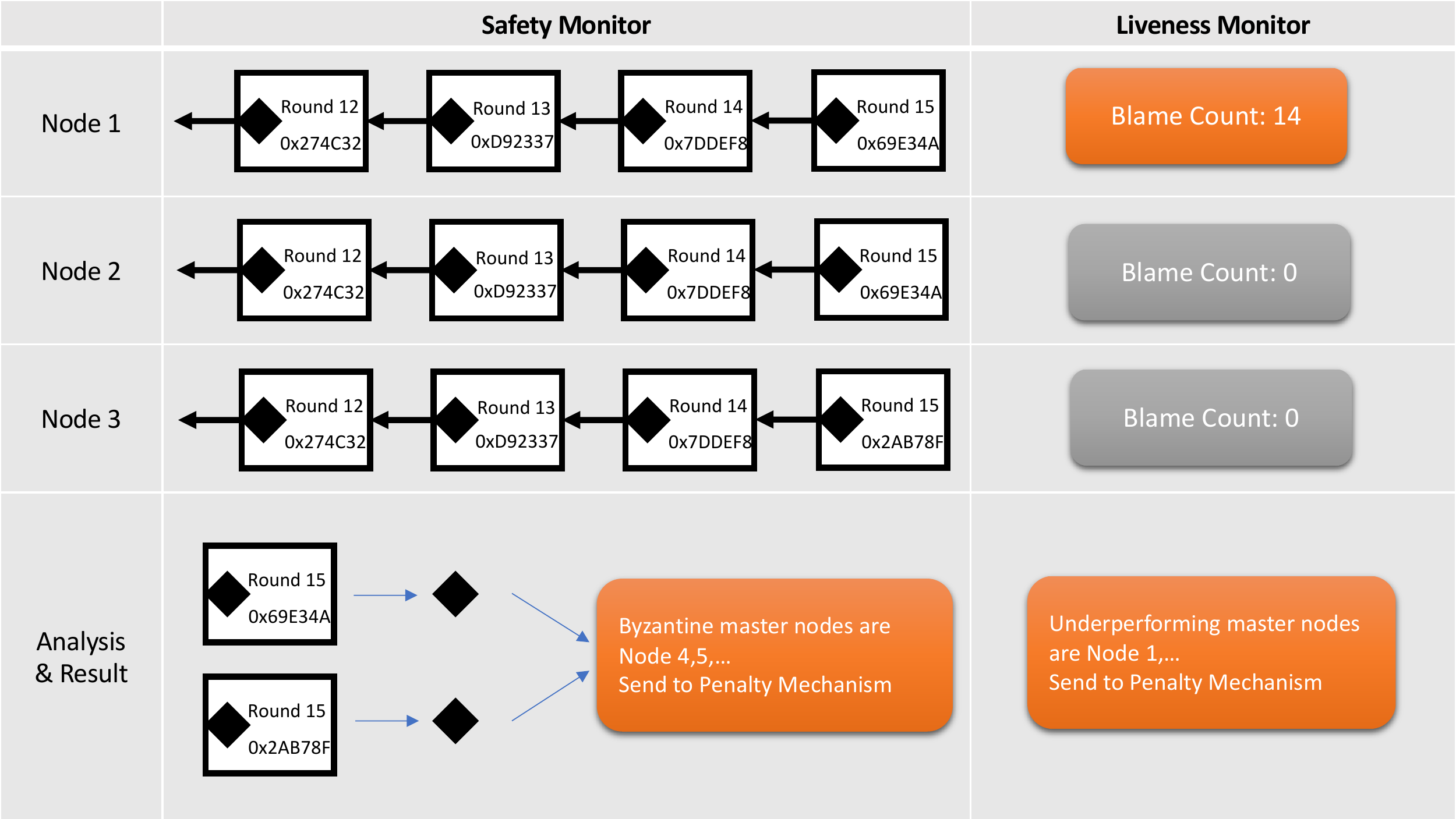}
    \caption{Detector dashboard.}
    \label{fig:demo}
\end{figure}
We will build a dashboard for the detectors to visualize the forensic information. Figure~\ref{fig:demo} is an artistic impression, which displays  information about each node's blockchain, blame messages, as well as safety and liveness forensic monitoring results.

%% file: section4.tex
\section{ Blockchains with Judiciary: An Outlook to the Future}

Over the past few years, speculation around crypto-currencies, their valuations and their utility has created a natural on ramp for crypto into the collective consciousness. While sentiment surrounding various blockchain ecosystems will remain volatile, there exists an ever-present gravitational pull towards community-driven adoption, which must be acknowledged and accepted. 

We are passing the point of diligent exploration and breaking through to stream entry, where widescale adoption of these blockchains ecosystems will be carried down the convincing currents of the mainstream. To prepare for a future of progressive adoption and a growing network effect, community-driven governance models must also emerge to complement expansion. This will ensure the crypto industry remains backed by secure, accountable networks, for the integrity of the system will depend on a balance of incentive-based and judiciary measures.

In the case of the XDC Network, which is positioned to disintermediate the trade finance industry, there is a pressing desire to create a security framework that emulates the accountability structure of correspondent banking systems.

Currently in trade finance, insufficient levels of collateral and complex regulatory and supply chain procedures create limitations for lenders. This increases the barrier to entry for institutions seeking to explore trade finance options, but more specifically, makes it difficult for MSMEs to find adequate funding. Due to the increasing costs of compliance and the volatility in the prices of commodities, there is an emerging pressure on this industry. 

For infrastructure development, especially in developing regions, projects need the enhanced visibility and competitive rates provided by a decentralized blockchain system with enterprise participants. Without these features, they will remain entrenched in the same systems that have long stifled their economic growth.  

We will call this forensic-enabled system Blockchain 4.0, for it will meet the unique needs of enterprises, by providing scalability, speed and systemic accountability. A necessary evolution to blockchain technology, XDPoS 2.0 will kick down the door to further enterprise adoption, which will enable competitive project funding and improved rates for SME participants. At this next stage, wide-scale protocol adoption will stem from enterprise participation in a trusted system, where institutional players maintain a level of control.

Conversely, the narrative propagating early Bitcoin adoption stemmed from libertarian ideals. Bitcoin, though embraced as a technological marvel in terms of security and decentralization, lacks the speed and scalability necessary to provide a suitable lending system for the trade finance industry. 

If blockchain is believed to displace the incumbent financial powers, and put power back into the hands of the people, a noble endeavor, how likely is this marketing campaign to encourage enterprise participation?

Under the current blockchain architecture, which lacks a judiciary system to ensure accountability, enterprises are asked to trade a material advantage, in this case security and control within their current system, in exchange for a solution deemed to be morally superior—an arrangement equated to a reverse Faustian bargain, unlikely to be accepted. 

There is little incentive for them to make such a trade, as enterprises have been rewarded for being a trusted party in the current system and could be punished by joining systems without accountable verifying parties, in this case validators.

Before institutions can make a substantial commitment to lend, borrow or transact on blockchain systems, they will ask two questions: 
(1)	What are the benefits?  
(2)	And how bad is the worst-case scenario?

Alternative lenders, specifically originators in the trade finance sector, therefore, stand to benefit from an answer that not only highlights the benefits, but also suggests in response to question 2: the worst-case scenario is not that bad. 

We, therefore, are challenged to create a psychological on-ramp for institutions, an on-ramp not buoyed by libertarian ideals, but rather reinforced by a promise of security, transparency, reduced counter-party risk, and cost-efficiency. Opportunely, these are all advantages inherent to the XDC Network. Throughout its established history, the network has clearly demonstrated while transactions are efficient and affordable, they are also immutable, secure, irreversible and permanently recorded. In fact, the XDC Network is already designed to withstand beyond a 51\% attack, raising the bar, substantially, to the extent that it would require control of over 75\% of network validators in order to successfully perform such an attack. However, there are other challenges within a Delegated Proof of Stake consensus mechanism that need to be addressed, such as ensuring security by being Byzantium Fault Tolerant.

Therefore, the evolution comes with a revolutionary forensics system, which could be equated to a ``new sheriff in town" in conjunction with an ``independent" judiciary system implemented via a forensics function, embedded into the network on a protocol level, which maintains an orderly ecosystem through systemic checks and balances, has the potential to reign in the wild west of DeFi, and reduce the time enterprises spend in the purgatory between indecision and adoption.

The new sheriff and judiciary system  would honor the moral ideals on which blockchain was founded, however, also provide a sufficient level of accountability and attributability, reliably course-correcting towards trustworthy cooperation. In the imminent future, DeFi arenas must remain well-patrolled by the forensic system we have outlined in the preceding sections. 

Accountability must cast a wide net and move away from centralized frameworks and control centers. Instead, it must be issued on a protocol level, by a decentralized ecosystem with living, breathing participants—a system in which all participants share equal, universal vulnerability. Among the XDC Network’s greatest strengths will be this decentralized forensics system, which is designed to issue accountability as unbiased as a hot stove to a child with a curious hand. 

The forensics system is a first step towards a vision where enterprises and retail users can have as fear-free an experience as tapping a credit card while buying groceries. 

A trade finance ecosystem with global reach, self-sustained record keeping, instant settlement and forensic monitoring is the perfect system to kick start the domino effect and force siloed monopolies on lending to compete with each other to move towards a more accommodating future. In an ecosystem that offers forensic security and equips institutions with the ability to create funding guarantees to back a trade, originators are also incentivized by reduced risk.

As blockchain networks of the future continue to rally enterprise interest and bring the promise of streamlined financing and supply chain processes, they also present the opportunity to identify and fill industry essential positions, reaching towards the libertarian ideals on which Bitcoin was initially marketed. Though it seems counterintuitive, a system designed for reducing corporate and institutional overhead, will likely enable workplaces to create retainable employment opportunities that will prove to be more secure in the coming years.

With these consensus engine enhancements, namely the advent of XDPoS 2.0, participants are ensured military grade security, augmented by forensic monitoring, which identifies misbehaving nodes. The network is backed by a perpetual witness, or all-seeing eye. In this system, bad actors are identified and penalty mechanisms are swiftly implemented. The network showcases a last line of defense, course-correcting mechanism to ensure cooperation and containment. Incorporating Byzantium Fault Tolerance onto a DPoS with forensic capability, coupled with the ability to shutdown malicious behavior establishes a new height for blockchains on which to aspire. 

The industry is in the progressive adoption phase, and the XDC network is prepared to lead the way with Blockchain 4.0, ensuring network transactions move like water through a stream, rather than a torch through a relay race, where any one party can fail. Thus, the XDC Network, implements this forensic feature to ensure seamless cooperation. XDPoS 2.0 inherently emulates the security of traditional trade finance procedures, with added accountability, on a protocol level, which enables the preservation of blockchain's premier advantages: the efficiency and cost advantages of a near instant settlement system; and it paves the way for enterprise adoption at scale.

%% file: acknowledgement.tex
\section{Acknowledgement}
The design of incorporating forensics in BFT protocols and the specific forensic protocol in the context of the XDPoS protocol are inspired by pioneering research work in {\sf Hotstuff} \cite{yin2019hotstuff} and ``BFT Protocol Forensics" \cite{sheng2020bft}.